\documentclass[aps]{revtex4}
\usepackage[]{graphicx}

\def\e{\mathop{\rm \mbox{{\Large e}}}\nolimits}

\newcommand{\be}{\begin{equation}}
\newcommand{\ee}{\end{equation}}
\newcommand{\bc}{\begin{center}}
\newcommand{\ec}{\end{center}}
\newcommand{\bea}{\begin{eqnarray}}
\newcommand{\eea}{\end{eqnarray}}
\newcommand{\ba}{\begin{array}}
\newcommand{\ea}{\end{array}}

\begin{document}

\title{AC-Stark effect in a semi-spherical quantum dot}

\author{Arezky H. Rodr\'{\i}guez$^{(a,b)}$, L. Meza-Montes$^{(a)}$,
C. Trallero-Giner$^{(b)}$ and S. E. Ulloa$^{(c)}$}

\affiliation{(a) Instituto de F\'{\i}sica, Universidad Aut\'onoma de Puebla,
Apdo. Postal J-48, Puebla, Pue. 72570, M\'exico. \\
(b) Departamento de F\'{\i}sica Te\'orica. Universidad de La Habana, Vedado 10400,
Cuba. \\
(c) Department of Physics and Astronomy, Condensed Matter and Surface Science
Program, Ohio University, Athens,Ohio 45701-2970}

\begin{abstract}
We present a theoretical approach to study the effects of an ac-field applied to
quantum dots with semi-spherical symmetry. Using the Floquet formalism for this
periodically driven system, the time-dependent Hamiltonian in the effective mass
approximation is solved. We show that the Hilbert space of solutions is separated
into orthogonal subspaces with different $z$-component of the angular momentum. We
give an explicit analytical representation for electronic states as a function of the
intensity and frequency of the electric field. Under the two level approximation, two
particular cases are studied: the low- and high-frequency regimes, which result of
comparing the ac-field frequency to the characteristic level splitting at zero field.

\end{abstract}

\maketitle                   

\section*{}

The dynamics of charged particles in semiconductor structures under a time-dependent
external electric field has been a topic of intense research
\cite{grifoni1998,ulloa2002,creffiedl2003,trallero2004,ulloa2004}. The progress of
techniques on the nanoscale has made possible the systematic study of effects only
present in intense alternating fields in this domain.

In this paper we address the problem of a quantum dot (QD) of semi-spherical shape
with the ac-field along the axial symmetry of the dot ($z$-axis in this case) in a
nonperturbative approach. To do so we make use of the Floquet formalism
\cite{grifoni1998}. The Hamiltonian for an electron in the QD under an electric field
of intensity $F$ and frequency $\omega$ can be written as
\be
H = H_o + |e| \, F \, z \, sin(\omega t),
\ee
where $H_o = - (\hbar^2 / 2 m^*) \nabla^2 + V_o$ is the Hamiltonian for the nondriven
system. Here $V_o = 0$ inside the QD and we use hard-wall boundary conditions, $m^*$
is the effective mass of the electron in the QD.

In the absence of the ac field the problem is exactly soluble and its eigenenergies
are given by the square of the zeroes $\mu_n^{(l)}$ of the spherical Bessel
functions, where $n$ is the order of the zero and $l$ the order of the Bessel
function $J_{l+1/2}$ (see details in \cite{prb}). The eigenfunctions, in spherical
coordinates, are given by
\be
\label{2}
f^{(o)}_{n,l,m}(r,\theta,\phi) = \frac{1}{\sqrt{r}} \frac{J_{l+1/2}(\mu_n^{(l)} \,
r)}{N_J} \frac{P_l^{|m|}(\cos\theta)}{N_P} \frac{\e^{\textstyle{i \, m\,
\phi}}}{\sqrt{2\pi}},
\ee
where $P_l^{|m|}$ are the associated Legendre polinomials, $m$ is the $z$-component
of the angular momentum, and $N_J$ and $N_P$ are normalization constants.

The dynamics of the driven system is governed by the time-dependent Schroedinger
equation
\be
\label{3}
i \hbar \frac{\partial \Psi}{\partial t} = H \Psi.
\ee

Since $H$ is periodic in time with period $\tau=2\pi/\omega$, it is possible to use
the standard Floquet theory \cite{grifoni1998} and write the wavefunction as
$\Psi=\exp(-i\,\varepsilon\,t/\hbar) \Phi$, which allows us to rewrite Eq. (\ref{3})
as
\be
\label{5}
\left( H - i\,\hbar \frac{\partial}{\partial t} \right) \Phi(r,\theta,\phi,t) =
\varepsilon \, \Phi(r,\theta,\phi,t),
\ee
where $\varepsilon$ is a real-valued parameter termed the Floquet characteristic
exponent, or the quasienergy.

To solve Eq. (\ref{5}), $\Phi$ can be expanded in a Fourier series for the temporal
component and linear combinations of stationary states
\be
\label{6}
\Phi(r,\theta,\phi,t) = \sum_{p=-\infty}^{\infty} \exp(i\,p\,\omega\,t) \;
u_p(r,\theta,\phi), \;\; p\in {\cal Z},
\ee
where the index $p$ labels the corresponding photon replicas \cite{ulloa2002}. The
functions $u_p$ are expanded in a series of solutions of $H_o$ given by Eq.
(\ref{2}). Due to the axial symmetry, the Hilbert space of solutions is separated
into orthogonal subspaces with different $z$-component of the angular momentum $m$.
In what follows, we consider the subspace defined by $m = 0$, therefore, the number
$m$ is omitted. Then, the expansion of $u_p$ is taken only over the pair of quantum
numbers ${n,l}$ which are labelled by the single index $\alpha$:
\be
\label{7}
u_p(r,\theta,\phi) = \sum_{\alpha=\{n,l\}}^\infty C_\alpha^{(\xi,\Omega)}(p)
f^{(o)}_{\alpha}.
\ee

We introduced the dimensionless quantities $\xi=F/F_o$ and $\Omega=\hbar\,\omega/E_o$
where the units of energy and ac field intensity are $E_o = \hbar^2 / 2 m^* a^2$ and
$F_o = E_o / |e| a$, respectively, being $a$ both the QD radius and the unit of
distance. Substitution of Eqs. (\ref{6}) and (\ref{7}) in (\ref{5}) leads to the
time-independent eigenvalue problem, in terms of an infinite set of equations
\be
\label{eigenvalue}
\sum_{I = \{\alpha,p\}} \left[ \left( \lambda - (\mu_{\alpha})^2 - p \; \Omega
\right) \delta_{I,I'} - \, \frac{\xi}{2} \; {\rm Z}(\alpha,\alpha') \;
(\delta_{p',p-1} + \delta_{p',p+1}) \right] C_I = 0,
\ee
where ${\rm Z(\alpha,\alpha')} = <f^{(0)}_\alpha |r\cos\theta|f^{(0)}_{\alpha'}>$ are
the matrix elements coupling zero-field states. The quasienergy spectrum is
determined by the interaction (coupling) among the nondriven levels and their
replicas. Thus, when only one nondriven state is considered (i.e., one value of
$\alpha$ in Eq. (\ref{eigenvalue})), the interplay between the replicas is such that
there is no variation of the levels at all as the electric field changes.

\begin{figure}
\includegraphics[width=.43\textwidth]{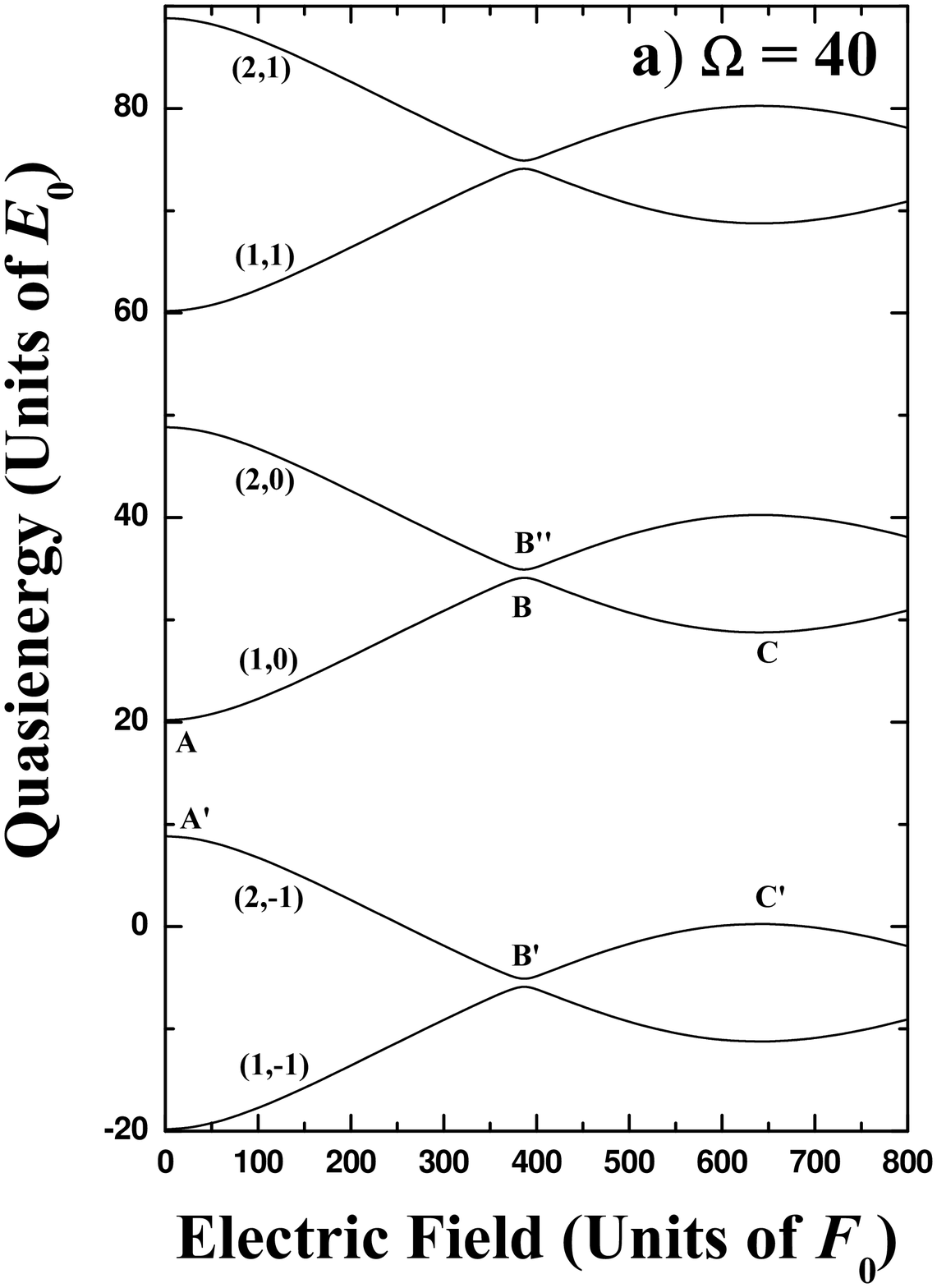}
\includegraphics[width=.43\textwidth]{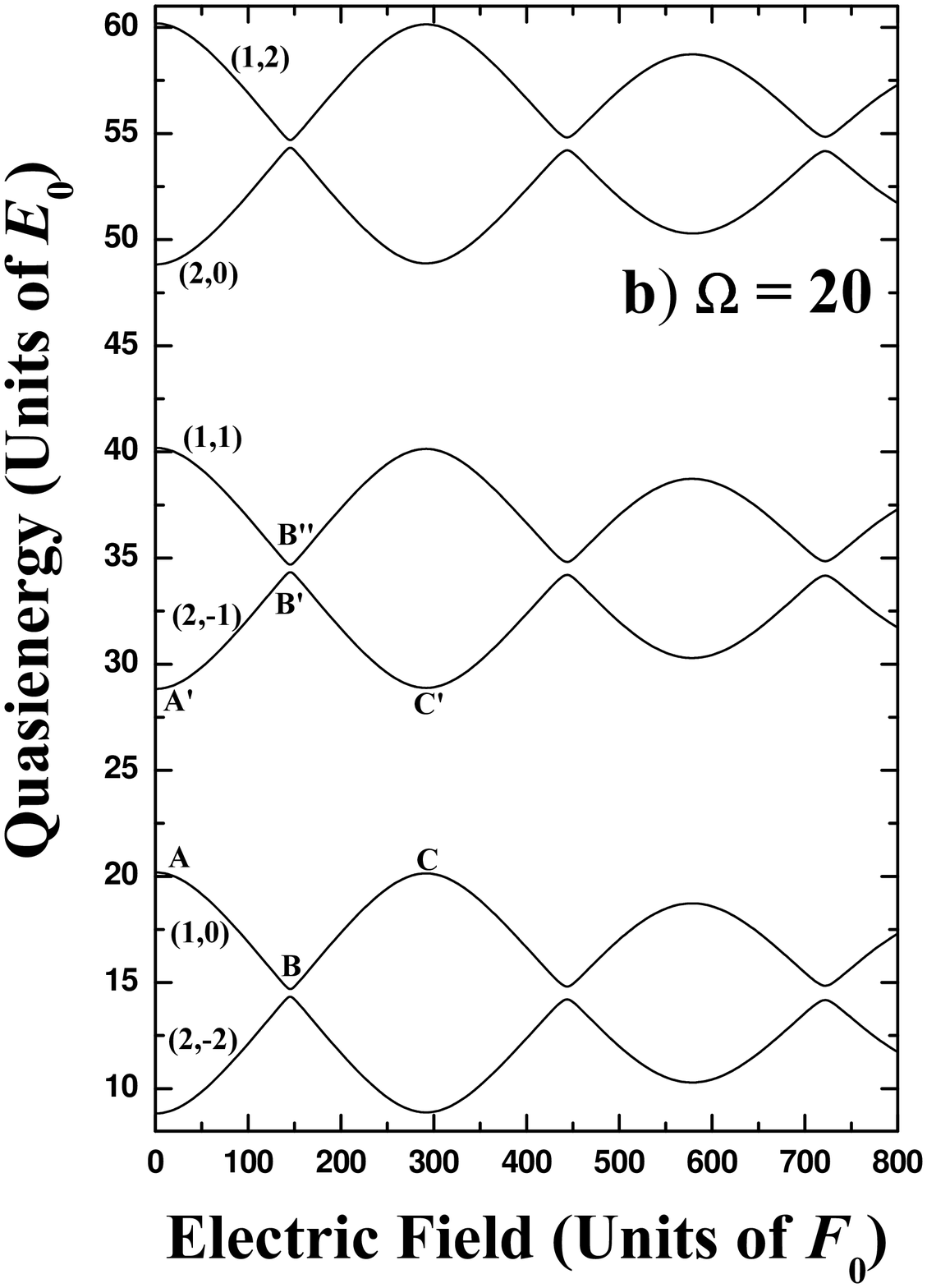}
\caption{Some quasienergies with $m=0$ as a function
of the field for two zero-field energy levels. The indexes ($\alpha,p$) are used to
indicate the corresponding quasienergy at zero field. a) High-frequency regime with
$\Omega = 40$, b) low-frequency regime with $\Omega = 20$.}
\label{fig1}
\end{figure}

In the framework of the two nondriven levels, which are the ground and the first
excited states at $m=0$, we obtain the quasienergy distributions shown in Fig.
\ref{fig1}. According to the chosen units, $800 F_0$ represents approximately 100
kV/cm for an InAs QD with $a = 50$ nm and for a CdSe QD with $a = 30$ nm. The field
intensity would increase as the radius diminishes. In Fig. \ref{fig1}a) we present
the high frequency regime ($\Omega
> \Delta$, where $\Delta \approx 28.7$ is the dimensionless gap between these two
levels) and in Fig. \ref{fig1}b) the low frequency regime ($\Omega < \Delta$). In
both cases, suitable ranges of quasienergies were chosen for the analysis. At $F=0$,
the pair of numbers ($\alpha,p$) are good quantum numbers, indicating the nondriven
level and the corresponding replica. At $F\neq 0$, the field introduces mixing
between states and $(\alpha,p)$ are not longer good quantum numbers to describe the
spectrum, (see Eq. (\ref{eigenvalue})). In Fig. \ref{fig1}a), the oscillating
behaviour of quasienergy (1,0) is explained as follows: as the second term in Eq.
(\ref{eigenvalue}) shows, this level only interacts with the nearby levels (2,-1) and
(2,1), where it holds the condition $\Delta p = \pm 1$ with $\Delta p = p' - p$. We
have not considered the interactions with their own replicas (1,-1) and (1,1) because
they are negligible compared to the interactions with the closest level (2,-1). The
proximity between levels (1,0) and (2,-1) makes them split off as the electric field
is turned on, leading to the anticrossing in points A and A'. The separation between
these levels can not increase unlimitedly because if it were greater than $\Omega$ it
would produce a degeneracy between, for example, levels (1,0) and (2,0), which is not
possible since they have the same spatial symmetry. On the other hand, the observed
repulsion between levels (1,0) and (2,0) at points B-B'' is mediated through the
coupling between levels (1,0) and (2,-1) and the periodicity of the system. As levels
(1,0) and (2,-1) start to approach each other again, their direct interaction gives
rise to the anticrossing C-C'. A similar explanation can be given for the replicas.

In the low-frequency regime shown in Fig. \ref{fig1}b), the anticrossings A-A',
B-B'-B'' and C-C' can be explained following the previous analysis. It can also be
seen that the anticrossings occur at smaller values of the field, compared to Fig.
\ref{fig1}a). To understand why, let us focus our attention on points A and A'. In
this case, the level (1,0) is now closer to level (2,-1) than in Fig. \ref{fig1}a)
and, in addition, interacts with level (2,1), which is not shown. While in Fig.
\ref{fig1}a), these two levels are down and above level (1,0), respectively, here
both of them are located above. Also notice that, in the high-frequency regime, the
distribution of nearby levels with respect to level (1,0) produces repulsion over
level (1,0) at opposite directions and in consequence a soft change as electric field
increases, whereas in Fig. \ref{fig1}b) both levels repeal it in the same direction,
causing a stronger effect. On the other hand, the more intense the electric field the
higher the weight of the matrix elements ${\rm Z}(\alpha,\alpha')$ in Eq.
(\ref{eigenvalue}) and, in consequence, stronger admixture of the levels widens the
gaps among them.

In conclusion, we have analyzed a semi-spherical quantum dot in the presence of an
oscillating electric field. We have found, in agreement with recent analytical work
in double and multiple quantum wells \cite{ulloa2002,ulloa2004}, that the electric
field at which the anticrossing occurs changes monotonically down as frequency
decreases. An explanation was given in terms of the interaction among zero-field
levels and their replicas. We have also studied the behaviour of the gaps between
quasienergies at the anticrossing points and have found that the gaps increase with
higher electric field due to the stronger admixture of the levels. When two
zero-field levels are taken into account, there are two types of anticrossings. One
of them (see points C and C' in each Figure) is due to a ``direct'' interaction
between two nearby levels with $|\Delta p| = 1$. This type is characterized by a
smooth variation of the quasienergy levels and occurs along a wide range of the field
intensity. The second one (see points B and B'' in each Figure) is due to an
``indirect'' interaction among the levels and the time periodicity of the system. It
is characterized by an abrupt change which occurs in a very narrow interval of the
field intensity. We also gave an explicit analytical representation for the functions
defined by Eq. (\ref{3}). These results can be very important and helpful for the
interpretation of experimental data or theoretical calculations for driven tunneling
structures in coupled multiple self-assembled quantum dots, and also to describe
optical properties in the presence of an external ac field. Moreover, they can also
be used to calculate the probability density and the time evolution of finding one
particle in each quasienergy state.

Partially supported by CONACyT - Proyecto E36764, Academia Mexicana de Ciencias and
Fundaci\'on M\'exico - Estados Unidos.

\end{document}